\documentclass[prb,twocolumn,unsortedaddress,amsmath,amssymb]{revtex4}

\usepackage{graphicx}
\usepackage{dcolumn}
\usepackage{bm}

\begin{document}

\preprint{ATB-1}

\title{Magnetization of La$_{2-x}$Sr$_x$NiO$_{4+\delta}$
 ($0 \leq x \leq 0.5$) and observation of novel memory effects.}

\author{P. G. Freeman}
\homepage{http://xray.physics.ox.ac.uk/Boothroyd}
\author{A. T. Boothroyd}
\author{D. Prabhakaran}
\affiliation{Department of Physics, Oxford University, Oxford, OX1
3PU, United Kingdom }

\author{J. Lorenzana}
\affiliation{SMC-INFM, ISC-CNR, Dipartimento di Fisica,
Universit\`{a} di Roma La Sapienza, P. Aldo Moro 2, 00185 Roma,
Italy}

\date{\today}

\begin{abstract}
We have studied the magnetization of a series of spin--charge
ordered La$_{2-x}$Sr$_x$NiO$_{4+ \delta}$ single crystals with
$0\leq x \leq$\ 0.5. For fields applied parallel to the $ab$ plane
there is a large irreversibility below a temperature $T_{\rm
F1}\sim 50$\,K and a smaller irreversibility that persists up to
near the charge ordering temperature. We observed a novel memory
effect in the thermo-remnant magnetization across the entire
doping range. We found that these materials retain a memory of the
temperature at which an external field was removed, and that there
is a pronounced increase in the thermo-remnant magnetization when
the system is warmed through a spin reorientation transition.

\end{abstract}

\maketitle

\section{\label{sec:intro}Introduction}

In the last decade it has become apparent that hole-doped
antiferromagnetic oxides have a strong tendency to form complex
ordered phases involving spin and charge degrees of freedom. Among
the most studied such materials are the layered superconducting
cuprates La$_{2-x}$Sr$_x$CuO$_4$(LSCO) and the iso-structural but
non-superconducting nickelates La$_{2-x}$Sr$_x$NiO$_4$(LSNO), both
of which exhibit a spin--charge ordered `stripe'
phase.\cite{tranquada-Nature-1995,chen-PRL-1993} Also common to the
phase diagram of both these systems is a so-called `spin glass'
phase, identified from irreversible behavior in magnetization
measurements.\cite{Chou-PRL-1995,spinglass}

The close proximity of the stripe and spin glass phases suggest that
these phenomena might be related. Evidence of the glassy nature of
stripe phases has already been found in neutron diffraction
measurements. Tranquada {\it et al.} studied the stripe-ordered
phase of La$_{1.6-x}$Nd$_{0.4}$Sr$_x$CuO$_4$ with $x = 0.12$ and $x
= 0.15$ and observed slowly fluctuating short-range magnetic
correlations which persisted above the bulk magnetic ordering
temperature indicating a glassy transition to the ordered
state.\cite{tranquada-PRB-1999} Similarly, diffraction measurements
on LSNO have shown that neither the charge ordering transition nor
the magnetic ordering transition are particularly well
defined.\cite{Du-PRL-2000,Hess-PRB-1999,Lee-PRL-1997,Yamamoto-PRL-1998,freeman-PRB-2002}
Further, a recent study of the bulk magnetization in superconducting
LSCO found that the irreversible  magnetization of LSCO behaves in a
manner that resembles the fundamental properties of the
superconducting state.\cite{panagopoulos-PRB-2004} The possibility
of a connection between stripes, spin glass behavior and
superconductivity is clearly of great interest.

The origin of the irreversibility in LSNO and LSCO is not well
understood. A straightforward explanation is that quenched
disorder frustrates the magnetic interactions and produces a
canonical spin-glass.\cite{spinglass} The puzzling results of
Ref.~\onlinecite{panagopoulos-PRB-2004} in LSCO, however, suggest
that superconducting correlations above $T_c$ may be responsible
for the irreversibility seen in that material.

The purpose of the present work was to investigate the
relationship between irreversibility effects observed in
magnetization measurements and the ordering properties of the
stripe phase. We chose to study the LSNO system because stripe
ordering extends over a wide range of doping and has been
characterized in
detail.\cite{chen-PRL-1993,yoshizawa-PRB-2000,lee-PRB-2001,kajimoto-PRB-2003,freeman-PRB-2004}
In addition, LSNO is not superconducting so, in principle, it
should help to distinguish irreversibility associated with charge
and spin ordering alone from irreversibility associated also with
superconductivity.

\begin{figure}[!ht]
\begin{center}
\includegraphics[width=8cm,clip=]{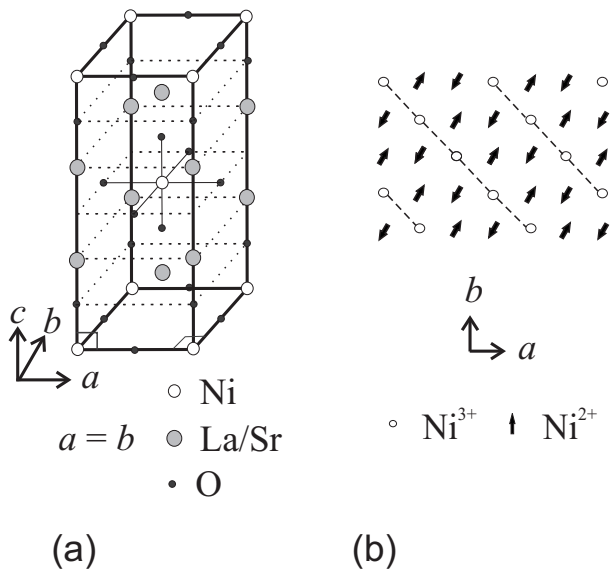}
\caption[Spin structure of La$_{5/3}$Sr$_{1/3}$NiO$_4$]{(a)
Tetragonal unit cell of La$_{2-x}$Sr$_x$NiO$_4$. (b) Pattern of
spin--charge ordering in the NiO$_2$ planes in
La$_{5/3}$Sr$_{1/3}$NiO$_4$. Arrows denote $S = 1$ spins on the
Ni$^{2+}$ ions and open circles represent holes, here assumed
centred on Ni sites. The broken lines indicate the charge stripes.
The O sites are not shown for clarity. }\label{Sr0.33spin}
\end{center}
\end{figure}

The basic stripe pattern in LSNO is illustrated in
Fig.~\ref{Sr0.33spin}.\cite{tranquada-PRL-1993} Holes introduced
into Mott-insulating NiO$_2$ layers by Sr or O doping arrange
themselves into an array of parallel lines in a background of
antiferromagnetically ordered Ni$^{2+}$ spins. The charge stripes,
which are aligned at $45^{\circ}$ to the Ni-O bonds, act as
anti-phase domain boundaries to the antiferromagnetic order. The
stripe ordering pattern that forms at $x = 1/3$, shown in
Fig.~\ref{Sr0.33spin}, is special because it is commensurate with
the underlying square lattice of the NiO$_2$ plane with both the
charge and magnetic order having the same period. This leads to a
particularly stable ordering at this doping
level.\cite{ramirez-PRL-1996,yoshizawa-PRB-2000,kajimoto-PRB-2001}
At other doping levels the spin and charge order are
incommensurate with the crystal lattice, and a model based on
periodically spaced discommensurations has been proposed to
explain the ordering wavevectors observed in diffraction
experiments. There is some evidence that the charge stripes are
located on the Ni sites, creating formally Ni$^{3+}$
ions,\cite{Li-PRB-2003} but there is also evidence that the holes
can reside at least for some of the time in oxygen
orbitals.\cite{Schuessler-Langeheine}, Spin degrees of freedom
also exist within the charge stripes but these do not form
long-range static magnetic order. There are, however,
quasi-one-dimensional antiferromagnetic correlations between the
spins in the charge stripes.\cite{boothroyd-PRL-2003}

The stripe-ordered phase has been observed in
La$_{2-x}$Sr$_{x}$NiO$_4$ for Sr doping in the range 0.135 $\leq x
\leq$ 0.5, and also in some oxygen-doped compounds
La$_{2}$NiO$_{4+\delta}$.\cite{Lee-PRL-1997,neutron,yoshizawa-PRB-2000,lee-PRB-2001,kajimoto-PRB-2001,kajimoto-PRB-2003,freeman-PRB-2002,freeman-PRB-2004,Du-PRL-2000,x-ray,pash-PRL-2000,hatton-2002,Ghazi-PRB-2004}
Charge ordering occurs at a temperature $T_{\rm CO}$ typically
100--200\,K depending on doping. Magnetic order occurs at a slightly
lower temperature $T_{\rm SO}$. An exception is $x=0.5$ which has an
anomalously high charge ordering temperature of $T_{\rm CO} \simeq
480$\,K due to the particular stability of the checkerboard charge
ordering pattern that forms at half
doping.\cite{chen-PRL-1993,kajimoto-PRB-2003} Below $T_{\rm IC}
\approx 180$\,K the checkerboard pattern becomes slightly
incommensurate, and below $T_{\rm SO} \simeq 80$\,K it is
accompanied by incommensurate magnetic order. The origin of these
incommensurate effects is not well understood. The correlation
lengths of the charge and stripe order in these materials are
typically $100-300$\AA\ for $0.2 < x \leq 0.47$ with both
correlation lengths being particular long in commensurately ordered
$x = 1/3$.\cite{yoshizawa-PRB-2000,Ghazi-PRB-2004}


An interesting feature of the magnetic order in LSNO is the
existence of a spin reorientation transition. This transition
features strongly in the present work. It was first observed in the
$x = 1/3$ and $x=1/2$ materials at a temperature $T_{\rm SR}$ of
50\,K and 57\,K, respectively\cite{lee-PRB-2001,freeman-PRB-2002} At
the spin reorientation transition the spins, which lie in the $ab$
plane at a non-trivial angle to the crystal axes, were observed to
rotate within the $ab$ plane through an angle of 13$^{\circ}$
($x=1/3$) or 26$^{\circ}$ ($x=1/2$).  Recently, a similar spin
reorientation was found in $x = 0.275, 0.37$ and $x = 0.4$
compositions but at a reduced temperature of $\sim
15$\,K.\cite{freeman-PRB-2004}


In this work we studied the magnetization of a series of LSNO single
crystals covering a wide range of doping levels. Our work
complements the recent analysis of the high temperature
magnetization described by Winkler {\it et
al.}.\cite{Winkler-PRB-2002} A preliminary account of some of our
results was given in Ref.\ \onlinecite{Freeman-JMMM-2004}, and here
we describe our experiments in more detail. We observe irreversible
magnetic behavior in all the samples for the case where the magnetic
field was applied parallel to the $ab$ plane. We also observed for
this field orientation some interesting memory effects
associated with the magnetic irreversibility.

\section{\label{sec:exper}Experimental Detail}

Single crystals of La$_{2-x}$Sr$_{x}$NiO$_{4+\delta}$ with $0\leq
x\leq 0.5$ were grown in Oxford by the floating-zone
technique.\cite{Prab} Typical dimensions of the crystals used in
this work were $\sim$$5\times5\times2$\,mm$^{3}$. The oxygen excess
$\delta$, determined by thermogravimetric analysis (TGA), is given
for each crystal in Table \ref{tab:deltaagain}. Our results for the
variation of $\delta$ with $x$ are broadly consistent with a
previous report for zone-melted crystals.\cite{Jang-JJAP-1991}
Crystals with $x \leq 0.25$ have significant excess oxygen, whereas
those with $0.275\leq x\leq 0.5$ are almost stoichiometric. The only
exception is the crystal with $x=0.2$ which has an anomalously low
value of $\delta$. No TGA measurement was carried on the $x = 0.37$
composition, but from Table \ref{tab:deltaagain} we would expect
$\delta \simeq 0.01$ since this crystal was grown under similar
conditions to the others. Moreover, the temperature dependence of
the magnetic and charge order of crystals from the same batches as
those studied here has previously been measured for most Sr
compositions by neutron or x-ray
diffraction.\cite{hatton-2002,Ghazi-PRB-2004,freeman-PRB-2002,freeman-PRB-2004}
The incommensurability of the $x = 0.37$ crystal was found to be in
line with that of the $x = 0.333$ and $x = 0.4$ crystals (see
Fig.~\ref{fig:ordedTs}).

\begin{table}
\begin{ruledtabular}
\begin{tabular}{ccc}
$x$ & $\delta\ (\pm 0.01)$ & $n_h$\\
\hline
\\
0&0.11&0.22\\
0.1&0.075&0.25\\
0.2&0.01&0.22\\
0.225&0.07&0.365\\
0.25&0.06&0.37\\
0.275&0.02&0.315\\
0.3&0.01&0.32\\
0.333&0.015&0.36\\
0.37& --- & ---\\
0.4&0.005&0.41\\
0.5&0.02&0.54\\
\hline
\end{tabular}
\end{ruledtabular}
\caption{Oxygen excess $\delta$ of crystals of
La$_{2-x}$Sr$_{x}$NiO$_{4+\delta}$ determined by thermogravimetric
analysis. The nominal hole content $n_h$ is given by $n_h = x +
2\delta$. \label{tab:deltaagain}}
\end{table}

Magnetization measurements were carried out with a SQUID
magnetometer (Quantum Design). The measurements were made by the
d.c. method, either with the magnetic field parallel to the $ab$
plane ($H \parallel ab$) or parallel to the crystal $c$ axis ($H
\parallel c$). For some of the measurements with $H \parallel c$ we used a
rotating sample mount to align the crystal accurately via the
anisotropy of the magnetization. The rotating mount contributes a
significant magnetic background which had to be measured and
subtracted from the signal. Any uncertainty in the subtraction will
cause a systematic error in the magnetization, and so measurements
taken with the rotating mount will be indicated.

Temperature scans of the magnetization were performed either by
measuring while cooling the sample in an applied field of 500\,Oe
(FC), or by cooling the sample in zero field and subsequently
measuring while warming in a field of 500\,Oe (ZFC). Typically the
data points were collected at a rate of one every $2-4$ minutes.
To study relaxation and memory effects we used several different
field/temperature protocols which will be described later.

\section{\label{sec:Results}Results}
\subsection{\label{sec:Mag mesurements}Magnetization vs temperature}

\begin{figure}[!ht]
\begin{center}
\includegraphics[clip=]{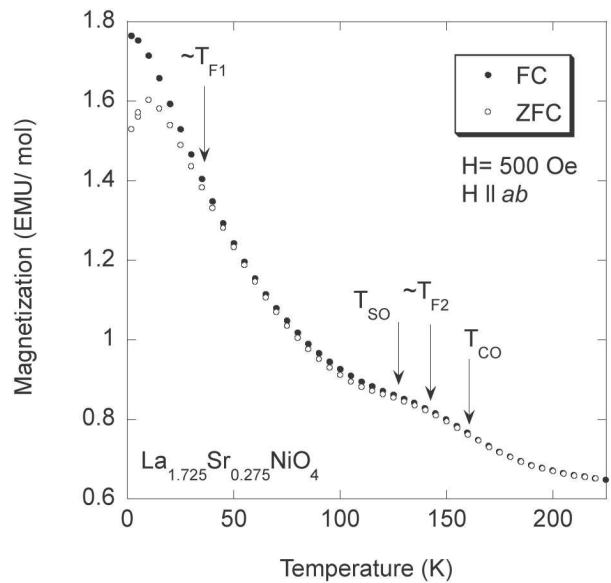}
\caption[Typical ZFC-FC data for LSNO]{Field-cooled (FC) and
zero-field-cooled (ZFC) d.c. magnetization of
La$_{1.725}$Sr$_{0.275}$NiO$_{4}$. The curves were measured with a
field of 500\,Oe applied parallel to the $ab$ plane. $T_{\rm SO}$
and $T_{\rm CO}$ are the magnetic and charge ordering temperatures,
respectively. The temperatures $T_{\rm F1}$ and $T_{\rm F2}$ are
explained in the text.} \label{fig:typical}
\end{center}
\end{figure}

Figure~\ref{fig:typical} shows a typical set of results for the FC
and ZFC magnetization of LSNO when the measuring field is applied
parallel to the $ab$ plane, in this case for a crystal with $x$ =
0.275. The FC and ZFC magnetization curves are seen to increase
with decreasing temperature and follow one another closely, except
at low temperature where there is a peak in the ZFC curve. We have
indicated approximate charge and magnetic ordering temperatures
$T_{\rm CO}$ and $T_{\rm SO}$ based on the temperature dependence
of the charge and magnetic superlattice Bragg peaks measured by
X-ray and neutron
diffraction.\cite{Ghazi-PRB-2004,freeman-PRB-2004} The charge and
spin correlations build up rather slowly, over several tens of
Kelvin, so the ordering temperatures are not precisely defined.
Nevertheless, charge ordering and magnetic ordering have quite
distinct effects on the magnetization, as can be seen in
Fig.~\ref{fig:typical}. On cooling below $T_{\rm CO}$ the
magnetization curve rises less steeply, but once the temperature
drops below $T_{\rm SO}$ the curve begins to rise more rapidly
again. This `wiggle' in the magnetization associated with $T_{\rm
CO}$ and $T_{\rm SO}$ is observed for all the samples with $x\geq
0.2$.

For most of the temperature range below $T_{\rm CO}$ the FC curve
lies above the ZFC curve, indicative of glassy behavior. The FC--ZFC
separation increases noticeably below the temperature marked $T_{\rm
F1}\approx 40$\,K which is somewhat higher than the temperature at
which the peak is observed in the ZFC curve. Above $T_{\rm F1}$ the
FC--ZFC separation is approximately constant up to a temperature
$T_{\rm F2}\sim T_{\rm CO}$, above which the FC and ZFC curves
become coincident.

\begin{figure*}[!ht]
\begin{center}
\includegraphics[width=12cm]{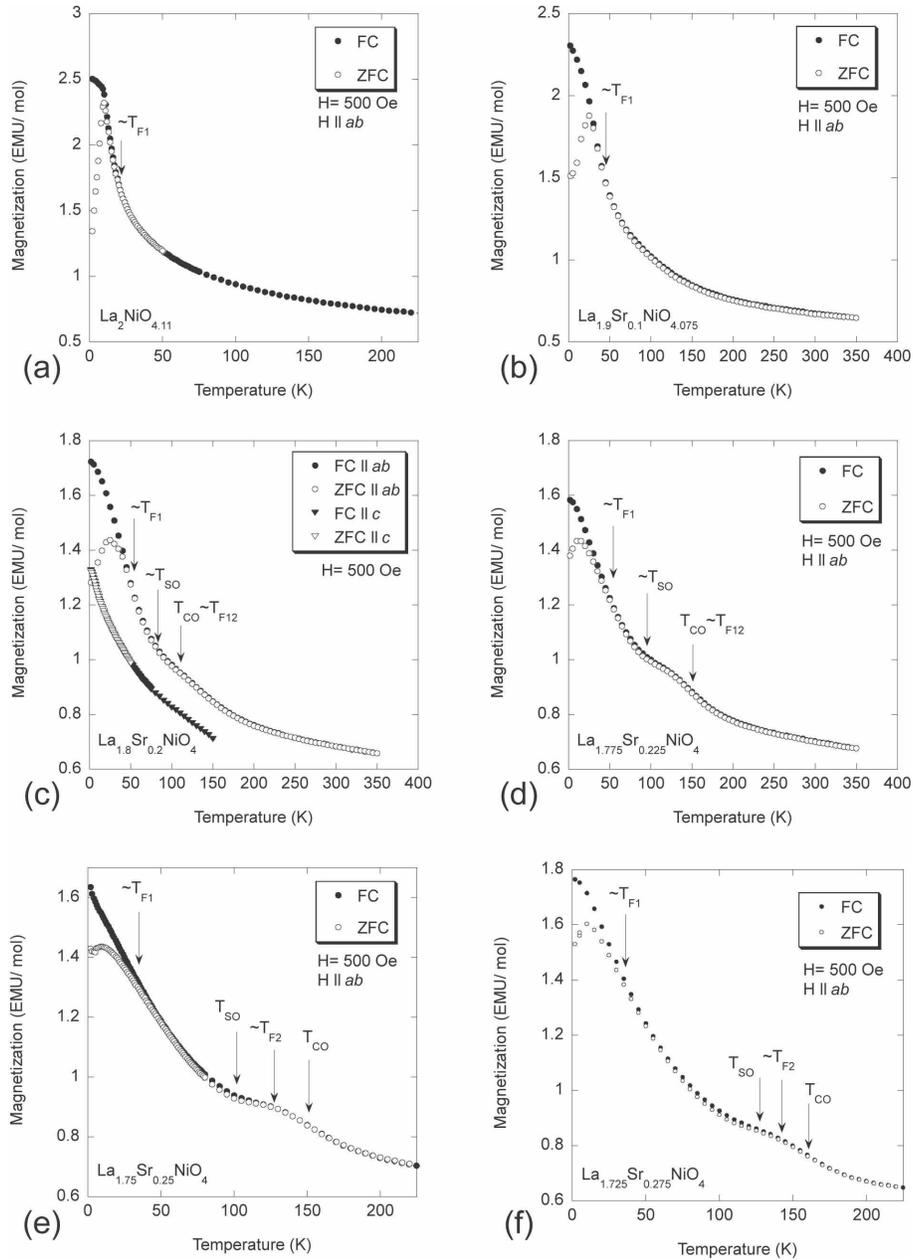}
\caption[ZFC-FC data for LSNO]{Field-cooled (FC) and
zero-field-cooled (ZFC) magnetization of single crystals of
La$_{2-x}$Sr$_{x}$NiO$_{4+\delta}$ with $0 \leq x \leq 0.275$. }
\label{fig:Spinglass1}
\end{center}
\end{figure*}

\begin{figure*}[!ht]
\begin{center}
\includegraphics[width=12cm]{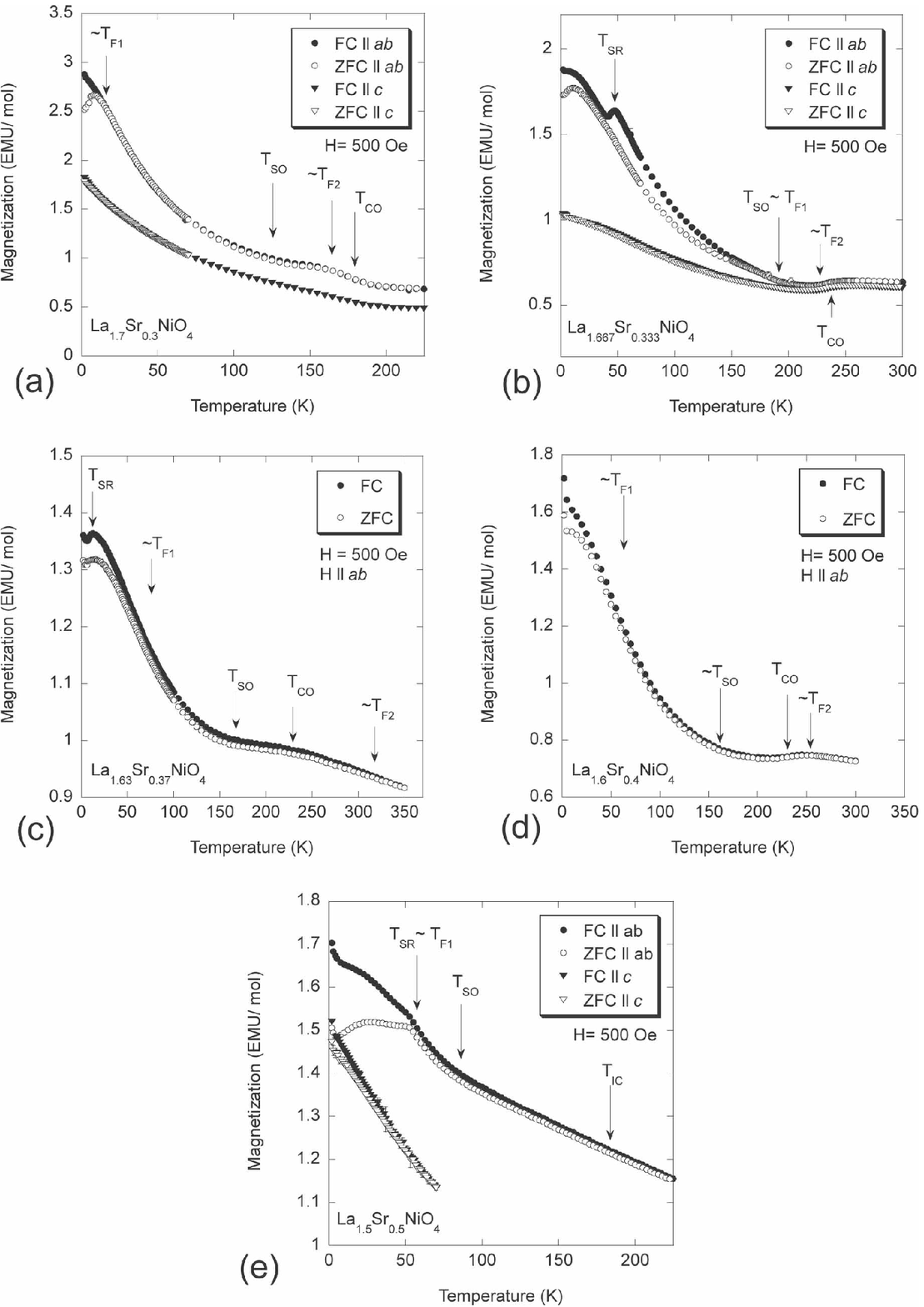}
\caption[ZFC-FC data for LSNO]{Field-cooled (FC) and
zero-field-cooled (ZFC) magnetization of single crystals of
La$_{2-x}$Sr$_{x}$NiO$_{4+\delta}$ with $0.3 \leq x \leq 0.5$. A
rotating sample mount was employed to obtain the data with
$H\parallel c$\ for the $x = 1/3$ crystal.} \label{fig:Spinglass2}
\end{center}
\end{figure*}

We carried out magnetization measurements on the eleven samples of
La$_{2-x}$Sr$_x$NiO$_{4+ \delta}$ listed in Table\
\ref{tab:deltaagain}. The results are presented in Figs.\
\ref{fig:Spinglass1} and \ref{fig:Spinglass2}. The curves with $H
\parallel ab$ all show similar features to those we have already
described for the $x=0.275$ sample. In some cases there is an
extra feature labeled $T_{\rm SR}$ associated with the spin
reorientation
transition.\cite{lee-PRB-2001,freeman-PRB-2002,freeman-PRB-2004}
This feature is especially prominent for the samples with  $x =
0.333$ ($T_{\rm SR}\simeq 50$\,K) and 0.5 ($T_{\rm SR}\simeq
57$\,K). For $x = 0.37$ and 0.4 (Fig.~\ref{fig:Spinglass2}) we
observe the temperature $T_{\rm F2}$ to be higher than $T_{\rm
CO}$, whereas for the other compositions apart from $x=0.5$ we
find $T_{\rm F2}$ to be similar to, or just below, $T_{\rm CO}$.
Our measurements on the $x=0.5$ sample did not extend high enough
to reach the charge ordering temperature of $T_{\rm CO} \simeq
480$\,K.

For several of the samples (those with $x=0.2$, 0.3, 0.333 and 0.5)
we measured magnetization curves with $H \parallel c$. These curves
are also presented in Figs. \ref{fig:Spinglass1} and
\ref{fig:Spinglass2}. In each case the curves with $H
\parallel c$ lie below those with $H \parallel ab$, consistent
with previous observations.\cite{Jang-JJAP-1991} Interestingly, when
$H \parallel c$ we observe little or no difference between the FC
and ZFC magnetization for these samples.

\begin{figure}[!ht]
\begin{center}
\includegraphics{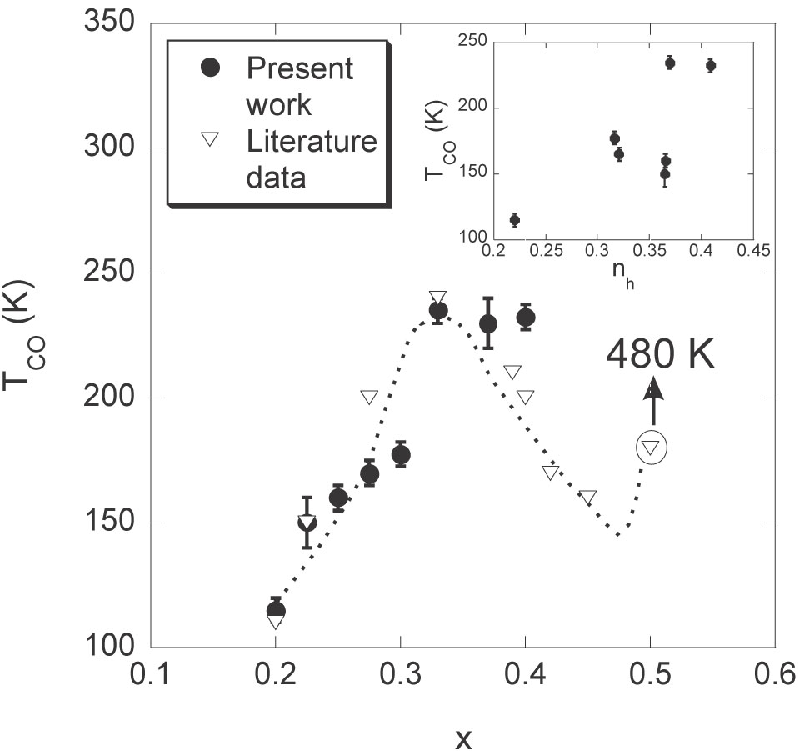}
\caption[Charge-order Temperatures for LSNO]{Charge ordering
temperature of single crystals of La$_{2-x}$Sr$_x$NiO$_{4+\delta}$
as a function of Sr doping $x$. The inset shows the charge ordering
temperatures plotted against the effective hole concentration $n_h =
x + 2\delta$. Literature data are taken from Refs.
\onlinecite{neutron,Lee-PRL-1997,yoshizawa-PRB-2000,lee-PRB-2001,kajimoto-PRB-2003,kajimoto-PRB-2001}.
} \label{fig:ordedTs}
\end{center}
\end{figure}

In Fig.\ \ref{fig:ordedTs} we plot the charge ordering temperatures
for our crystals determined either directly by X-ray diffraction or
from the wiggle in the magnetization.  For comparison we have
included results for LSNO published by other
groups.\cite{neutron,Lee-PRL-1997,yoshizawa-PRB-2000,lee-PRB-2001,kajimoto-PRB-2003,kajimoto-PRB-2001}
We have not shown any data for our $x=0$ and $x=0.1$ crystals
because although these are expected to exhibit charge ordering based
on the total hole count $n_h = x + 2\delta$ we did not examine these
crystals by X-ray or neutron diffraction and there is no feature in
the magnetization data in Figs.\ \ref{fig:Spinglass1}(a) and (b)
that we can identify with a charge ordering transition. Our results
are reasonably consistent with the literature results. In the
literature such phase diagrams are often plotted as a function of
$n_h$ rather than $x$ on the assumption that as far as the spin and
charge ordering temperatures are concerned oxygen doping is
equivalent to Sr doping. This assumption does not appear to be valid
for our samples. The inset to Fig. \ref{fig:ordedTs} showing $T_{\rm
CO}$ against $n_h$ for our samples does not have a smooth variation,
whereas the main plot of $T_{\rm CO}$ against $x$ does. An
inequivalence between Sr doping and oxygen doping was also found in
the magnetic ordering temperatures of a large set of polycrystalline
samples of LSNO by Jest\"{a}dt {\it et al.} using
$\mu$SR.\cite{Jestadt-PRB-1999}



\begin{figure}[!ht]
\begin{center}
\includegraphics{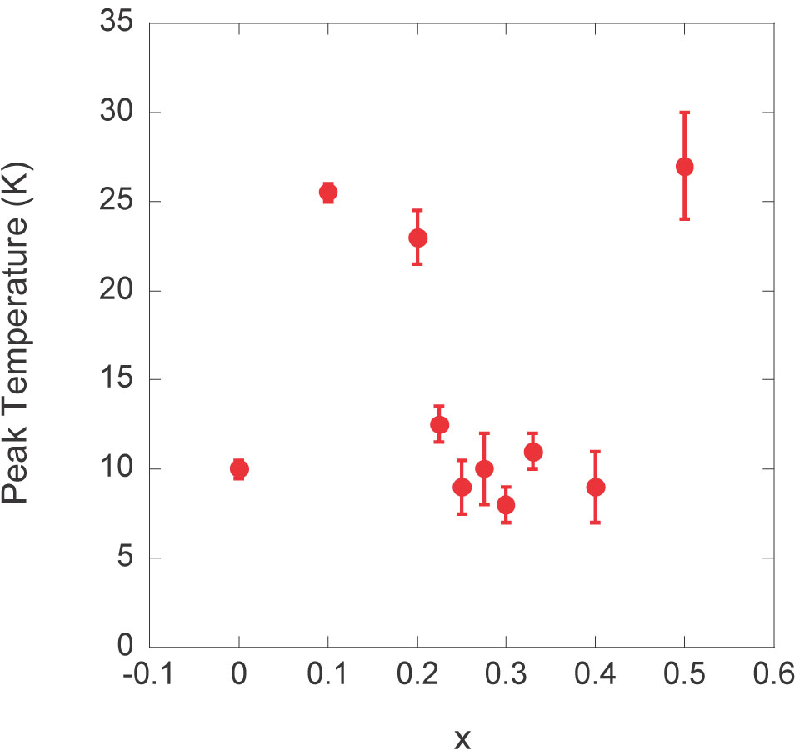}
\caption[Characteristic Temperatures of the Spin Glass State]{
Variation with Sr doping of the temperature at the peak of the ZFC
magnetization of La$_{2-x}$Sr$_x$NiO$_{4+\delta}$. The points are
obtained from the data shown in Figs.\ \ref{fig:Spinglass1} and
\ref{fig:Spinglass2}} \label{fig:SpinglassTs}
\end{center}
\end{figure}

In figure \ref{fig:SpinglassTs} we record the temperature of the ZFC
magnetization peak for each of the samples. This peak temperature is
seen to occur at $\sim10$\,K for most doping levels. For $x = 0.1$,
0.2 and $0.5$ the peak is at a somewhat higher temperature for
reasons that are unclear.

\begin{figure}[!ht]
\begin{center}
\includegraphics{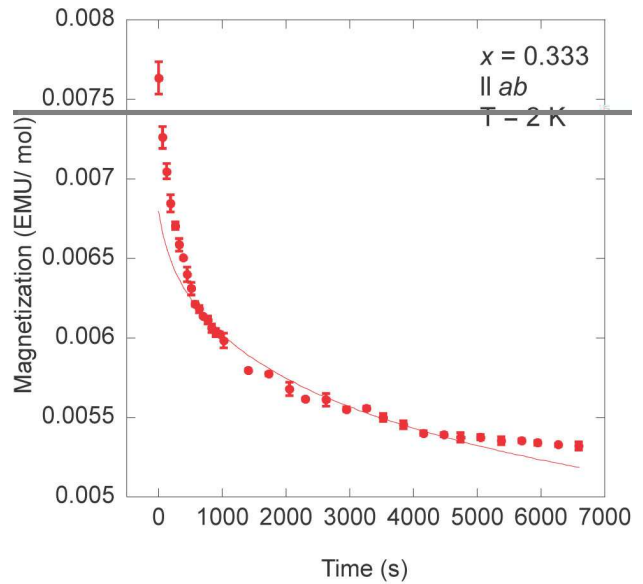}
\caption[Remnant induced signal in $x = 1/3$]{(Color online) Time
dependence of the remnant magnetization for a crystal of
La$_{2-x}$Sr$_x$NiO$_4$ with $x = 0.333$. The signal was induced by
application of a magnetic field of strength 500\,Oe parallel to the
$ab$ plane for 5 minutes at $T = 2$\,K. The time is measured after
the field has been switched off. The curve is the best fit to a
stretched exponential.} \label{fig:plop}
\end{center}
\end{figure}

The observation of features in the magnetization of stripe-ordered
LSNO indicative of spin glass and spin freezing behavior (the
FC--ZFC difference and the peak in the ZFC magnetization) suggests
that the system is out of thermodynamic equilibrium at low
temperatures and that relaxation effects in the experimental time
scale may be important. In fact, a time-dependent remnant signal was
reported some time ago by Lander {\it et al.} for a crystal with
$x=0.15$.\cite{Lander-PRB-1991} To investigate this effect in more
detail we performed the following experiment. We applied a field of
500\,Oe parallel to the $ab$ plane at fixed temperature for 5
minutes, turned the field off, and measured the remnant
magnetization as a function of time. Figure \ref{fig:plop} shows the
results for the $x = 1/3$ crystal measured at 2\,K. The remnant
signal has an initial rapid decay on a time scale of $\sim$1000\,s
followed by a much slower decay extending beyond the duration of our
experiment ($\sim$7000\,s). We attempted to fit the remnant signal
with a stretched exponential $M(t) = M \exp\{ -\alpha t^{(1-n)}\}$,
as found to describe the magnetization in spin
glasses,\cite{Continentino-PRB-1986} but the quality of the best fit
is not satisfactory as can be seen in Fig.\ \ref{fig:plop}. The
remnant magnetization of Fig. \ref{fig:plop} is observed parallel to
the $ab$ across the doping levels studied in this work for fixed
temperatures below $T_{F2}$.
%



\subsection{\label{sec:Memory effects}Memory effects}

The measurements presented so far indicate that the response of
stripe-ordered LSNO to a magnetic field is partly irreversible, in
the sense that application of a field followed by cooling to low
temperatures creates a different state to cooling in zero field
followed by application of a field, at least on the timescale of the
measurement. As shown in Fig.\ \ref{fig:plop}, the magnetization is
time-dependent but the system does not reach a steady state at low
temperature even after several hours.

In this section we describe a novel memory effect associated with
the slow relaxation of the remnant magnetization. The first results
we describe were obtained with the following field--temperature
protocol: the sample is cooled from 300\,K to a temperature $T_{\rm
0}$ in a field of $500$\,Oe applied parallel to the $ab$ plane. Once
at $T_{\rm 0}$ the field is removed, the sample is cooled to 2\,K in
zero field, and the thermo-remnant magnetization (TRM) is measured
while warming the sample up through $T_{\rm 0}$.

\begin{figure}[!ht]
\begin{center}
\includegraphics{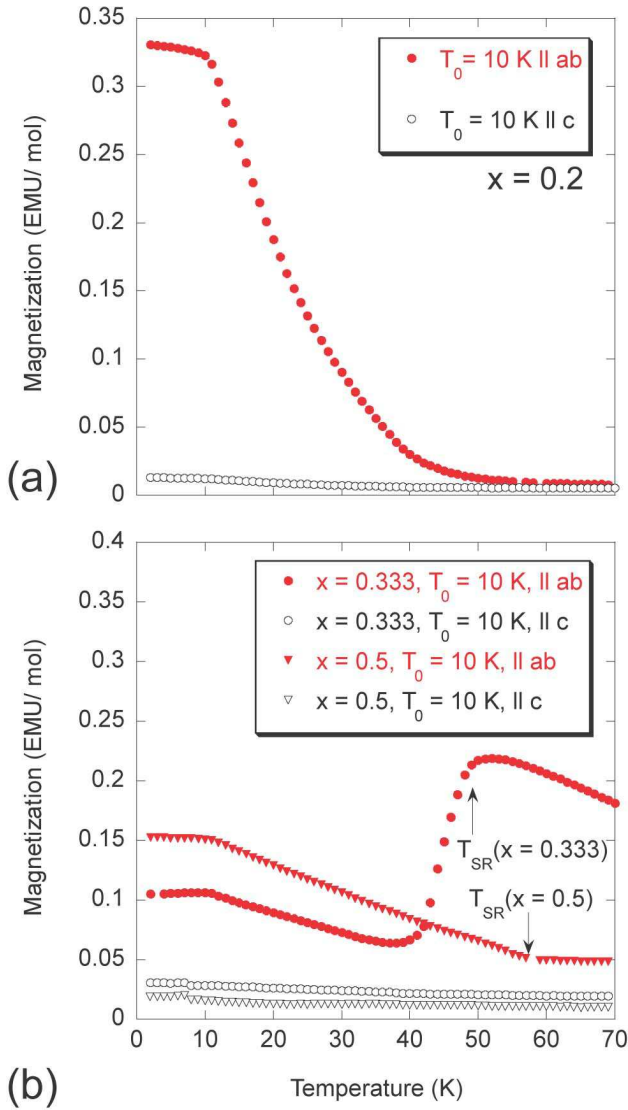}
\caption[Signal induced by FCing $\| c$]{(Color online) The
thermo-remnant magnetization (TRM) of La$_{2-x}$Sr$_x$NiO$_4$
induced by field-cooling down to $T_{\rm 0}= 10$\,K in a field of
$500$\,Oe followed by cooling to 2\,K in zero field and measuring
while warming in zero field. The symbols distinguish measurements
made with the field applied parallel to the $ab$ plane (closed
symbols) and parallel to the $c$ axis (open symbols). Panel (a)
shows data for $x = 0.2$, and panel (b) shows the corresponding
measurements for the $x=0.333$ and $x=0.5$ samples showing the
striking effects associated with the spin reorientation transition
(indicated by $T_{\rm SR}$) in these two compositions. The rotating
mount was used to obtain $H
\parallel c$ data for the $x = 0.333$ sample. } \label{fig:8}
\end{center}
\end{figure}

Figure \ref{fig:8}(a) shows a typical TRM response obtained with
this protocol. In this instance the data were obtained from the
$x=0.2$ sample with $T_{\rm 0} = 10$\,K. On warming from 2\,K the
TRM is fairly constant up to $T_{\rm 0}$, but above $T_{\rm 0}$ it
falls dramatically with increasing temperature. Hence, the system
has a clear `memory' of the temperature at which the field was
switched off.

The memory effect shown in Fig.\ \ref{fig:8}(a) was observed for
all the samples studied providing $T_{\rm 0}$ did not exceed
$T_{\rm F2}$. However, this is not the only interesting feature
revealed by this measurement protocol. Fig.\ \ref{fig:8}(b) shows
the TRM of samples with $x=0.333$ and $x=0.5$ using the same
protocol, again with $T_{\rm 0} = 10$\,K. The memory effect at
$T_{\rm 0}$ can clearly be seen, but there is also another
dramatic feature in the curves, this time at a temperature in the
region of 50\,K. For $x=0.333$ this feature takes the form of a
sharp upwards step in the curve, whereas for $x=0.5$ it is an
abrupt change in slope. This second feature is not present in the
data for $x=0.2$ shown in Fig.\ \ref{fig:8}(a), and we associate
it with the spin reorientations found near 50\,K in $x=0.333$ and
$x=0.5$ samples\cite{lee-PRB-2001,freeman-PRB-2002} (the $x=0.2$
sample does not have a spin reorientation near 50\,K).

\begin{figure}[!ht]
\begin{center}
\includegraphics{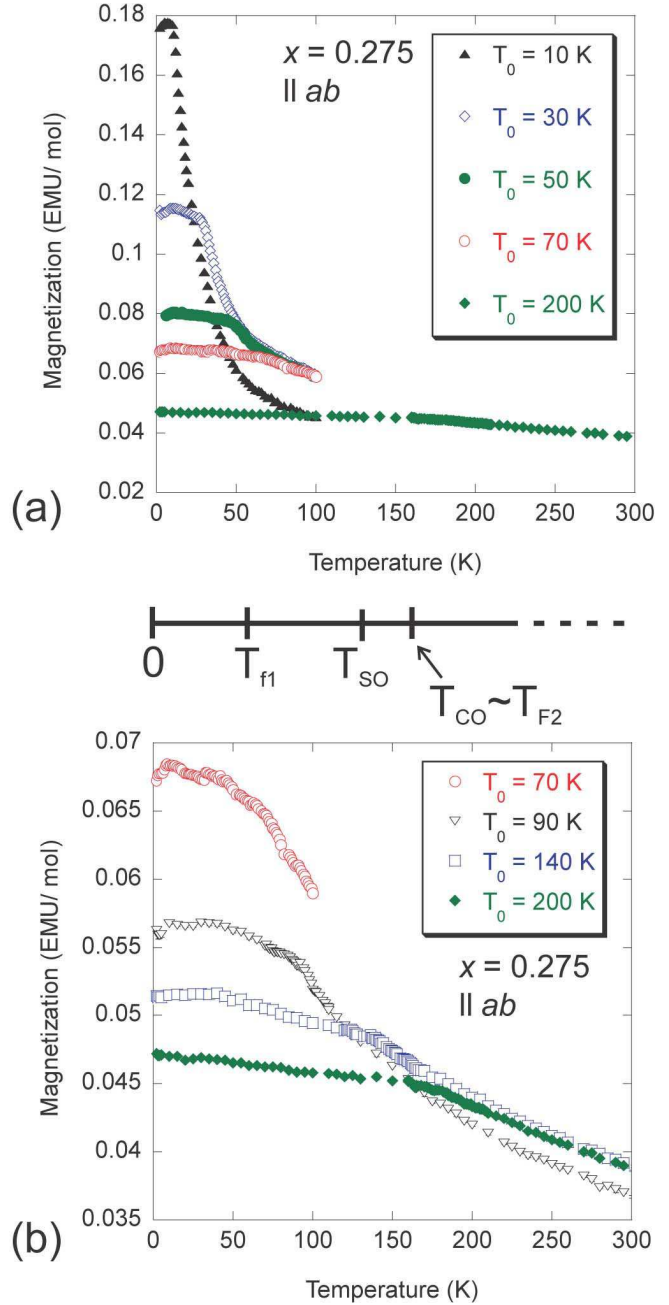}
\caption[Signal induced by FCing in $x = 0275$]{ (Color online) The
TRM of LSNO, $x=0.275$, for different values of $T_{\rm 0}$, where
$T_{\rm 0}$ is the temperature at which the magnetic field was
switched off in the protocol described in the text and in the
caption to Fig.\ \ref{fig:8}. (a) shows data for $T_{\rm 0} =
10$\,K, 30\,K, 50\,K, 70\,K and 200\,K, and (b) shows data for
$T_{\rm 0} = 70$\,K, 90\,K, 140\,K and 200\,K.}
\label{fig:FCing0_275}
\end{center}
\end{figure}

In Figs.~\ref{fig:FCing0_275}(a) and (b) we show results for the
$x=0.275$ crystal for several different $T_{\rm 0}$ values. As
$T_{\rm 0}$ increases the low temperature TRM systematically
decreases.
For $T>T_{\rm 0}$ the curves tend to fall one on top of each other.
Failure of the curve with $T_{\rm 0}=10$K to do so
could be due to fluctuations in the small residual field ($\sim$ few Oe) in
the magnet.

The TRM signal continues to decay above the charge
ordering temperature $T_{\rm CO}\simeq 160$\,K. The anomaly at
$T_{\rm 0}$ seems to be clearest when $T_{\rm 0} < T_{\rm F1}$
(i.e.\ the runs with $T_{\rm 0}= 10$\,K, 30\,K and 50\,K), but is
still discernible in the runs with $T_{\rm 0}= 70$\,K, 90\,K and
140\,K. There is no $T_{\rm 0}$ anomaly in the curve for $T_{\rm 0}=
200$\,K, but there is a change in slope in the region of $T_{\rm CO}$.
The $x=0.275$ composition is known to undergo a spin reorientation
below $T_{\rm SR}\simeq 12$\,K\cite{freeman-PRB-2004} and this may
explain the initial slight increase in the TRM up to $\sim T_{\rm
SR}$.

\begin{figure}[!ht]
\begin{center}
\includegraphics{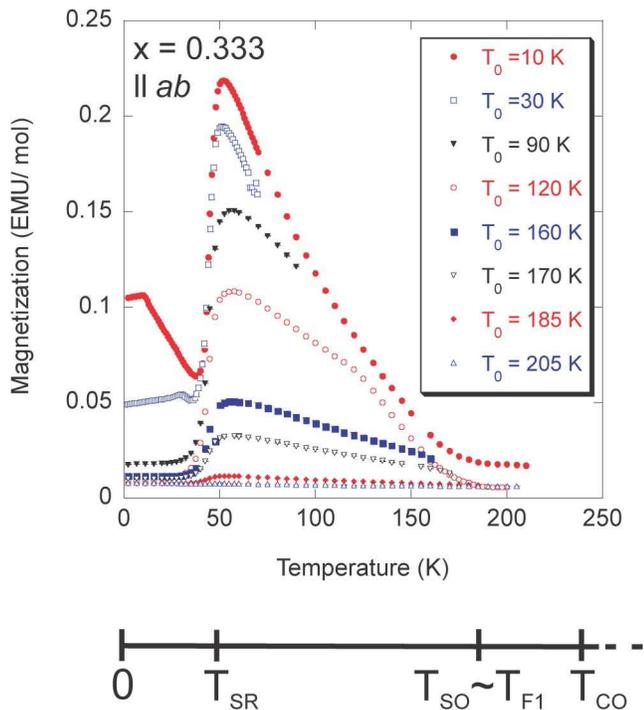}
\caption[Signal induced by FCing in $x = 1/3$]{(Color online) The TRM
of LSNO, $x=0.333$, for different values of $T_{\rm 0}$, where
$T_{\rm 0}$ is the temperature at which the magnetic field was
switched off in the protocol described in the text and in the
caption to Fig.\ \ref{fig:8}.} \label{fig:FCing0_333}
\end{center}
\end{figure}

Figure \ref{fig:FCing0_333} displays the TRM of the $x=0.333$
sample for several different $T_{\rm 0}$ temperatures. All the
curves apart from that for $T_{\rm 0}=205$\,K rise sharply to a
peak at a temperature close to $T_{\rm SR} \simeq 50$\,K. The
smaller the value of $T_{\rm 0}$ the larger is the peak. The
memory effect at $T_{\rm 0}$ is also present in these data, both
when $T_{\rm 0} < T_{\rm SR}$ and when $T_{\rm 0} > T_{\rm SR}$
(see for example the curve for $T_{\rm 0}=120$\,K) but not when
$T_{\rm 0}
> T_{\rm F1}$ (e.g.\ $T_{\rm 0}=205$\,K).


The memory effects at $T_{\rm 0}$ and $T_{\rm SR}$ just described
are only observed when the field is applied parallel to the $ab$
plane. As shown in Fig.\ \ref{fig:8}, for fields parallel to the $c$
axis the TRM is very small and although there is a hint of a $T_{\rm
0}$ anomaly in the $x=0.2$ and $x=0.5$ curves this could equally be
the result of a misalignment of the $c$ axis by a few degrees.

For some of the samples we investigated the dependence of the TRM on
several other parameters.  We found that the remnant magnetization
increased in size almost linearly with the inducing field, with
little indication of saturation for inducing fields up to 5\,T. We
also found that the magnitude of the induced signal was the same for
field-cooling rates from room temperature to $T_{\rm 0}$ of
$10$\,K/min and $3$\,K/min.

\begin{figure}[!ht]
\begin{center}
\includegraphics{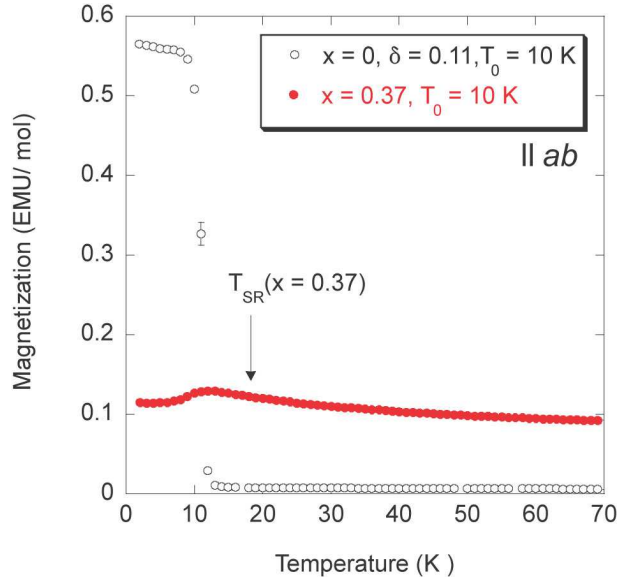}
\caption[Doping dependence of the signal induced by Fcing]{(Color
online)
The TRM of samples of LSNO with $x=0$, $\delta=0.11$ and
$x=0.37$. Here, $T_{\rm 0}=10$\,K is the temperature at which the
magnetic field was switched off in the protocol described in the
text and in the caption to Fig.~\ref{fig:8}. The spin-reorientation
temperatures $T_{\rm SR}$ of the $x = 0.37$ sample is indicated.}
\label{fig:inducedFC}
\end{center}
\end{figure}

Finally, Fig.\ \ref{fig:inducedFC} shows the TRM measured with the
usual field--temperature protocol for two other LSNO samples. Both
measurements were made with $T_{\rm 0} = 10$\,K. The $T_{\rm 0}$
anomaly is particularly strong in the sample with $x=0$ and
$\delta=0.11$. This may be because for this sample $T_{\rm 0}$
coincides with a very sharp peak in the ZFC magnetization [Fig.\
\ref{fig:Spinglass1}(a)]. For the $x=0.37$ sample the signal
around $T_{\rm 0}$ does not fall sharply like it does in the other
samples, but this is probably because the $T_{\rm 0}$ anomaly is
almost coincident with the memory signal associated with the spin
reorientation which is known to occur below $T_{\rm SR} = 19$\,K
in this sample.\cite{freeman-PRB-2004}


\section{\label{sec: Magnetism Discussion}Discussion and Conclusions}

The magnetization curves for the La$_{2-x}$Sr$_x$NiO$_{4+\delta}$
compounds reported here show a number of common features. For fields
applied parallel to the $ab$ plane there is a large irreversibility
below a temperature $T_{\rm F1}$ and a smaller irreversibility that
persists up to $T_{\rm F2}\sim T_{\rm CO}$ (Figs.\
\ref{fig:Spinglass1} and \ref{fig:Spinglass2}). A peak is observed
in the ZFC magnetization at temperatures in the range 10--30\,K.
This peak is generally rather rounded, reminiscent of a spin-glass
freezing transition. For the sample with $x=0$, $\delta = 0.11$ the
peak is much sharper, more like a transition to long-range magnetic
order except that the neutron diffraction study of Nakajima {\it et
al.} on a crystal of the same composition found a transition to
long-range antiferromagnetic order at $T_{\rm N}\simeq 50$\,K rather
than at 10\,K.\cite{Nakajima-JPSJ-1997} The magnetization curves
also exhibit features associated with spin reorientation
transitions, identified here with the help of prior neutron
diffraction results. These features are particularly prominent for
the $x=0.333$ and $x=0.5$ samples, but have a different effect on
the magnetization. For the $x=0.333$ sample there is a large FC--ZFC
separation above $T_{\rm SR}$, whereas for the $x=0.5$ sample a
large FC--ZFC separation occurs below $T_{\rm SR}$. It is not clear
why the two samples behave differently, especially given that the
nature of the spin reorientation is believed to be the same in both
cases (only the angle through which the spins turn is
different).\cite{lee-PRB-2001,freeman-PRB-2002}

The fact that the irreversibility in the magnetization is only
observed for fields parallel to the $ab$ plane implies that it
derives from the spins in the antiferromagnetic regions between the
charge stripes (see Fig.\ \ref{Sr0.33spin}). These spins have a
small XY-like anisotropy, and in the magnetically ordered phase they
line up parallel the $ab$ plane. It is then reasonable to assume
that the irreversibility originates from some degree of disorder in
the array of stripes, the disorder being either quenched or
self-generated. Schmalian and Wolynes\cite{Schmalian-PRL-2000} have
shown that stripe systems with competing interactions on different
length scales (here the short-range magnetic exchange and long-range
Coulomb interactions) can undergo a self-generated glass transition
caused by the frustrated nature of the interactions.
Also since the stripes are charged, quenched disorder due to the dopants
will frustrate the ideal periodic ordering of the stripes and produce
a complex energy landscape.
This stripe
glass state would have a large number of metastable states separated
by energy barriers.
Slow relaxation between these states would lead to relaxation behavior,
consistent with what we have observed here (Fig.~\ref{fig:plop}).

Physically, disorder in the stripe phase could take a number of
different forms. One source is magnetic frustration where a stripe
ends. From diffraction measurements it is known that the stripes
have a finite length.\cite{yoshizawa-PRB-2000,hatton-2002} At the
end of a charge stripe there is magnetic frustration where two
antiphase spin domains meet without the charged wall to stabilize
the antiphase configuration. Another possibility is variations in
the direction of the ordered magnetic moments. As noted above,
there is magnetic irreversibility associated with the spin
reorientation transitions that occur in some, if not all, striped
LSNO compounds. The ordered moments lie in the $ab$ plane, but in
general do not point along a symmetry direction within the
plane.\cite{freeman-PRB-2004} The system could therefore contain
spatially separated domains in which the moments point along
different equivalent directions and which could be unequally
populated. There is, in addition, some experimental evidence for a
distribution of moment directions,\cite{yoshinari-PRL-1999} which
could lead to disorder. A third possibility is that the finite
sized stripe domains could carry a net moment, and there could be
frustrated free spins at boundaries between the stripe domains.
Finally, there are the spin degrees of freedom within the charge
stripes to consider. Because these are at antiphase boundaries the
mean field coupling to the antiferromagnetic order is frustrated
(Fig.\ \ref{Sr0.33spin}). Although these spins do not exhibit
static long-range order there is evidence for short-range dynamic
antiferromagnetic correlations among
them,\cite{boothroyd-PRL-2003} and these correlations could at
least in principle freeze into a glassy state at low temperatures.
However, the dynamical susceptibility of these fluctuating spins
is observed to be largest in the $c$
direction,\cite{boothroyd-PRL-2003} and this is inconsistent with
the magnetization effects described here which only occur when the
field is parallel to the $ab$ plane.

Let us now discuss the memory effects observed here. Ageing and
memory effects are typical characteristics of spin
glasses,\cite{Jonason-PRL-1998} but here we have used a new protocol
and observed a phenomenologically different memory effect. We have
found that stripe-ordered nickelates have a memory of the
temperature at which an external field is removed, and also have a
memory of the state of the system at the spin reorientation
transition.


Ageing and memory effects in spin glasses can be understood
qualitatively in terms of jumps between a large number of metastable
states separated by energy barriers. Consider the decay in the
remnant magnetization on removing the magnetic field at constant
temperature shown in Fig.~\ref{fig:plop}. After the field is
switched off there is an initial phase in which the magnetization
decreases rapidly, with the system crossing small energy barriers
reversibly, but soon the system reaches a second phase in which the
magnetization decays much more slowly. During this second phase, the
system evolves via a series of ``quakes", i.e.\ large, irreversible,
configurational rearrangements involving many spins, driven by
extremal thermal fluctuations.\cite{Sibani-JSTAT-2004} Each quake
drives
a region of
the system from one metastable configuration into another one
with lower energy. The time needed to overcome a barrier of size
$\Delta$ is of order:
\begin{equation}\label{eq:t}
t(\Delta) \sim \tau_0 \exp\left( \frac{\Delta}{k_B T}\right)
\end{equation}
with $\tau_0$ a microscopic time. In a glassy state barriers are
broadly distributed. Barriers which satisfy $t(\Delta)\lesssim
t_{\rm exp}$, will lead to relaxation in the experimental time
scale $t_{\rm exp}$. However a significant fraction of the
barriers will satisfy $t(\Delta)> t_{\rm exp}$ leading to slow or
virtually zero relaxation in the experimental time scale. In
addition as the temperature is lowered it is expected that the
height of the barriers grows.\cite{bou02}
Hence, the decay in the magnetization eventually slows to a virtual
standstill before the true equilibrium ground state can ever be
reached.

Now consider the memory effect shown in Fig.~\ref{fig:8}(a). After
the field is switched off at $T_{\rm 0}$ the induced magnetization
decays as just described to a long-time metastable state
where barriers with $t(\Delta)> t_{\rm exp}$ hold parts of the
system with a non-zero out-of-equilibrium magnetization.
 On cooling, the amount of thermal energy available decreases
 and the barriers increase
 trapping the system in the
long-time state established at $T_{\rm 0}$. On re-heating the sample
the thermal fluctuations are insufficient to quake the system out of
its deep energy minimum as long as the temperature remains below
$T_{\rm 0}$. However, as soon as the temperature exceeds $T_{\rm 0}$
the decrease in the heights of the barriers and the increase on the
the thermal fluctuations allow new parts of the system to relax,
decreasing the TRM further.

\begin{figure}[!ht]
\begin{center}
\includegraphics{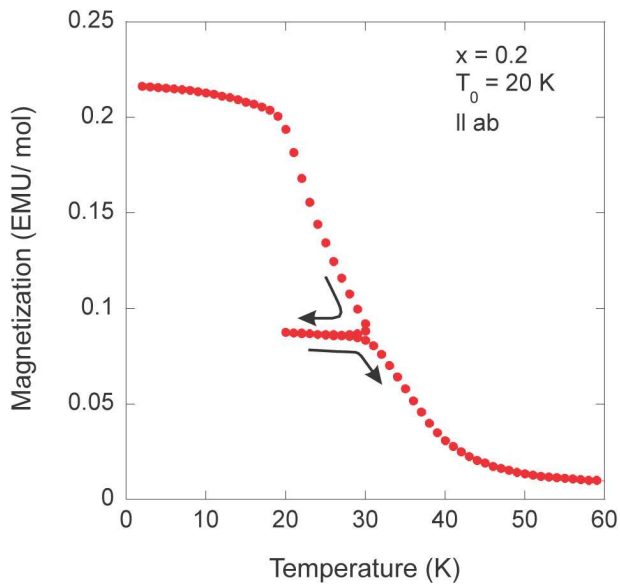}
\caption[Simple test]{(Color online) The TRM signal for the LSNO
sample with $x = 0.2$ resulting from the following protocol. First,
the sample was cooled from room temperature to $T_{\rm 0} = 20$\,K
in a field of $500$\,Oe applied parallel to the $ab$ plane, the
field removed and the sample cooled to 2\,K in zero field. The TRM
was then measured while warming from 2\,K to 30\,K, then while
cooling from 30\,K to 20\,K, and finally while warming from 20\,K to
60\,K.} \label{fig:test}
\end{center}
\end{figure}

If this description is valid then it should in theory be possible to
perform a cooling/re-heating excursion anywhere on the TRM curve at
$T>T_{\rm 0}$ and after the excursion return to the same TRM curve.
An experiment to test this prediction is presented in
Fig.~\ref{fig:test}. The $x=0.2$ sample was used, and we followed
the usual field--temperature protocol with $T_{\rm 0} = 20$\,K. As
expected, the TRM shows an abrupt drop on warming through $T_{\rm 0}
= 20$\,K. On reaching 30\,K, however, we stopped warming, cooled
back down to 20\,K and started warming again at the same rate,
measuring the TRM continuously through 30\,K up to 60\,K. As can be
seen in Fig.\ \ref{fig:test}, during the temperature excursion from
30\,K to 20\,K and then back again to 30\,K the TRM remains almost
constant. On further warming the TRM returns to the original curve.
This behavior is consistent with the picture we have described.

One of our comments about Fig.\ \ref{fig:FCing0_275} was that the
lower the temperature $T_{\rm 0}$ the larger is the TRM. This is
hardly surprising. First, the magnetization induced by the applied
field increases with decreasing temperature
(due, it must be
assumed, to the existence of effective free spins associated with
disorder),
and second, at lower $T_{\rm 0}$ the thermal
fluctuations are smaller and the barriers are higher
so more regions of the system contribute to the TRM.
 As already mentioned, we observe a large TRM for
$T_{\rm 0} < T_{\rm F1}$ and a small TRM for $T_{\rm F1} < T_{\rm
0} < T_{\rm F2} \sim T_{\rm CO}$. This is evidence that the cause
of the TRM is the same as that of the irreversible magnetization
in the FC-ZFC protocol.

Perhaps the most dramatic effect we have observed is the
remarkable increase in the TRM at the spin reorientation
transition of the $x = 0.333$ sample (see Fig.~\ref{fig:FCing0_333}).
Usually, TRM effects are characterized by a
reduction in remnant magnetization with increasing temperature,
whereas here we observe an increase of up to one order of
magnitude. This is especially surprising given that the FC
magnetization exhibits only a small drop on cooling through
$T_{\rm SR}$ [Fig.\ \ref{fig:Spinglass2}(b)].
 Memory effects have
been observed in a number of different systems,\cite{sasaki} but
we are not aware of any other compound that exhibits a memory
effect associated with a spin reorientation transition.
To some extent the memory effects reported here are simpler than the
phenomena reported in other glassy system in that our protocol does
not involve a waiting time where the system ages but are only
dependent on the cooling and heating protocol at relatively
fast rates.

In summary, we have observed irreversible behavior and memory
effects in the magnetization of La$_{2-x}$Sr$_{x}$NiO$_{4+\delta}$
when the magnetic field is applied parallel to the $ab$ plane. We
found particularly striking memory effects associated with a spin
reorientation transition. These observations suggest that
stripe-ordered LSNO has non-trivial dynamics, and it would be of
interest to find out if similar effects were present in other
stripe-ordered systems.


\section{\label{sec:Ack}Acknowledgements}

We would like to acknowledge P. Isla  and D. Gonz$\acute{a}$lez of
Instituto de Ciencia de Materials de Arag\'{o}n, CSIC-Universidad
de Zaragoza, Spain, for carrying out the thermogravimetric
analysis of the crystals used in this work. We would also like to
acknowledge that this work was supported in part by the
Engineering and Physical Sciences Research Council of Great
Britain. J.L. is in debt with F. Ricci-Tersenghi for enlightening
discussions on memory effects.

\end{document}